\documentclass{revtex4}

\usepackage{graphicx}
\usepackage{bm}
\usepackage{amssymb}
\usepackage{amsmath}
\usepackage{dcolumn}
\usepackage[latin1]{inputenc}
\usepackage[english]{babel}

\begin{document}


\title{Quasi Kepler's third law for quantum many-body systems}

\author{Claude \surname{Semay}}
\email[E-mail: ]{claude.semay@umons.ac.be}
\thanks{ORCiD: 0000-0001-6841-9850}

\author{Cintia T. \surname{Willemyns}}
\email[E-mail: ]{cintia.willemyns@umons.ac.be}
\thanks{ORCiD: 0000-0001-8114-0061}

\affiliation{Service de Physique Nucl\'{e}aire et Subnucl\'{e}aire,
Universit\'{e} de Mons,
UMONS Research Institute for Complex Systems,
Place du Parc 20, 7000 Mons, Belgium}
\date{\today}

\begin{abstract}
\textbf{Abstract} 

The Kepler's third law is a relation between the period and the energy of two classical particles interacting via a gravitational potential. Recent works showed that this law could be extended, at least approximately, to classical three-body systems, or even many-body classical systems. So, a classical quasi Kepler's third law seems to exist. In this paper, approximate analytical solutions are computed for quantum self-gravitating particles with different masses. The results give strong indications in favor of the existence of a quasi Kepler's third law for such systems. The relevance of the proposal is checked with accurate numerical data for the ground state of self-gravitating identical bosons and with numerical estimations for systems with identical particles plus a different one. Connections between the quantum and classical systems are discussed. 
\end{abstract}

\maketitle

\section{Introduction}
\label{sec:intro}

Kepler's third law played an important role in the history of physics. It recently reappeared in the spotlight when several accurate numerical computations showed strong indications that a generalized Kepler's third law exists for the classical motion of periodic three-body system \cite{dmit15,li17,li18,li19}. For classical two-body systems with a mechanical energy $E$ and a period $T$, the Kepler's third law states that  
\begin{equation}
\label{kepler3L}
T\,|E|^{3/2} =\frac{\pi}{\sqrt{2}}\,G\,\sqrt{\frac{(m_1\,m_2)^3}{m_1+m_2}},
\end{equation}
where $G$ is the gravitational constant, and $m_1$ and $m_2$ are their masses. Let us call this quantity the Kepler invariant $\tau$. For a classical three-body system, the definition of this quantity is given by
\begin{equation}
\label{tau3}
\tau = T^* |E|^{3/2},
\end{equation}
where $T^*=T/L_f$ takes into account the topology of the orbit around the three two-body collision points with the positive integer $L_f$. This number, which is named the ``free group element" of the orbit, is a complicated notion described in \cite{suva14}. To our knowledge, it is only defined for three-body systems. Because $\tau$ is found by numerical computation to be approximately equal to a universal constant for three bodies with the same mass, the existence of a quasi Kepler's third law can be proposed for such gravitational systems. Its extension for $N$-body periodic orbits is then questionable. Such a generalization has been proposed in \cite{sun18} and commented in \cite{zhao18}, based on dimensional analysis. This generalised law has been determined in order to coincide with the usual Kepler's third law for $N=2$ and to be in agreement with numerical results for $N=3$.

Strong connections exist between classical and quantum theories, the most famous one being certainly the Ehrenfest theorem, showing that expectation values obey Newton's second law. So, one can ask if Kepler's third law can also be
relevant for quantum $N$-body systems. This problem has been addressed in \cite{sema19a} for systems with identical particles. A Kepler invariant has also been found, but different from the one proposed for classical $N$-body systems. Within the quantum calculations, the periodic orbit is replaced by a stationary quantum state, and a quantum definition of the period must be used. A first definition of this period is proposed in \cite{sema19a} on the basis of a semiclassical approximation, but a more relevant definition (but numerically identical to the previous one) is given in \cite{sema19c}. In the following, the quantum Kepler invariant is computed (approximately but analytically) for a general many-body system and found to be identical to the proposal made in \cite{sun19}, using again arguments based on dimensional analysis.

It is worth mentioning that our calculation is not performed to bring relevant information about the links between gravitation and quantum mechanics. Indeed, there is no hope to experimentally check the validity of our result, due to the incredible weakness of the gravitational force. Our purpose is to explore the dynamics of many-body quantum systems and show that something special seems to happen with an inverse-square law force. 

The quantum Kepler invariant is defined in Sect.~\ref{sec:tauq}. The case of identical particles is already treated in \cite{sema19a}, but is presented in Sect.~\ref{sec:ident} for completeness, and because its relevance is checked with a particular example. Results for a general system are computed in Sect.~\ref{sec:gen}, where the relevance of the analysis is checked on a particular system composed of a set identical particles plus a different one. Some concluding remarks and outlooks are given in Sect.~\ref{sec:conclu}.

\section{Quantum Kepler invariant}
\label{sec:tauq}

The Hamiltonian for a system of self-gravitating particles is given by
\begin{equation}
\label{H}
H = \sum_{i=1}^N \frac{\bm p_i^2}{2\, m_i} - \sum_{i<j=2}^N \frac{G\, m_i\, m_j}{|\bm r_{ij}|}.
\end{equation}
The quantum Kepler invariant $\tau_q$ is computed with a stationary eigenstate of this Hamiltonian by the formula 
\begin{equation}
\label{tauq1}
\tau_q = T_q\, |E|^{3/2},
\end{equation} 
where $E = \langle H \rangle$ is the corresponding eigenvalue and $T_q$ the equivalent period, that is to say the equivalent of a period for a stationary quantum system \cite{sema19c}. $T_q$ is build from a classical definition for a period in which classical quantities are replaced by their quantum equivalent. It is given by
\begin{equation}
\label{Tq}
T_q = \frac{\pi\, I}{\langle K \rangle},
\end{equation}
where $I$ is the action for the state and $\langle K \rangle$ the mean value of $K$, the kinetic part of $H$. The action $I$ is easy to define for a particle \cite{sema19c}. But it is more tricky to compute for a many-body system. A proposal is made in \cite{sema19c}. The quantum many-body virial theorem \cite{luch90,ipek16,sema20} applied to (\ref{H}) implies that 
\begin{equation}
\label{virial}
E = -\langle K \rangle =\frac{\langle V \rangle}{2},
\end{equation}
where $V$ is the potential part of $H$. Actually, (\ref{virial}) is still valid if $(-G\, m_i\, m_j)$ is replaced by $k_{ij}$. The $k_{ij}$ can be positive or negative, provided a bound state exists. This equality can be checked by using the accurate numerical results from \cite{horn14} about the ground state of ``self gravitating bosons" and ``two-component Coulombic systems". To be fair, two results from the last examples have a relative error around 10\%, instead of less than 1\% for all other results. We did not identify the origin of these discrepancies. Using~(\ref{Tq}) and (\ref{virial}), (\ref{tauq1}) reduces to
\begin{equation}
\label{tauq}
\tau_q = \pi\, I\,|E|^{1/2}.
\end{equation}
The procedures to compute $I$ and $E$ are explained in the following sections.

\section{Quantum systems with identical particles}
\label{sec:ident}

The envelope theory (ET) \cite{hall80,hall81,hall83,hall95,hall04,hall19}, also known as the auxiliary field method \cite{silv10,sema13}, allows the solution of $N$-body quantum systems, eigenvalues and eigenvectors. In favorable situations, analytical upper or lower bounds can be obtained. The method is simple to implement and can provide fairly reliable results \cite{sema15a,sema15b,sema19b}. Most of the tests have been performed with the ground state, but the spectra can be calculated as well. An example is given in \cite{sema15a} for a three-quark system for which an average accuracy of a few percent has been reached for the 16 lowest levels. For the Hamiltonian~(\ref{H}) with identical particles ($m_i=m, \ \forall\ i$), the ET gives the following result \cite{sema15a}
\begin{equation}
\label{Emequal}
E_\textrm{id} = - \frac{N^2(N-1)^3}{16} \frac{G^2\, m^5}{Q_\phi(N)^2\,\hbar^2}, 
\end{equation}
where $Q_\phi(N)$ is a global quantum number given by
\begin{equation}
\label{Qphi}
Q_\phi(N)= \sum^{N-1}_{i=1} (\phi\, n_i+l_i) + (N-1) \frac{1+\phi}{2},
\end{equation}
$n_i$ and $l_i$ being the usual radial and orbital quantum numbers. Only specific values of $Q_\phi(N)$ are allowed, following the nature of the particles, bosons or fermions. In the original method, $\phi=2$ in (\ref{Qphi}) because the ET relies on the solutions of a many-body harmonic oscillator Hamiltonian \cite{silv10,sema13}. In the case $\phi = 2$, it can be shown that $E_\textrm{id}$ are upper bounds. By using the ET in combination with a generalization of the dominantly orbital state method, it is shown in \cite{sema15b} that the parameter $\phi$ can be introduced in the original global quantum number to improve the approximate energies. The idea comes from a result obtained in \cite{loba09} showing the existence of a universal effective quantum number for centrally symmetric 2-body systems. But the variational character of the ET can then not be guaranteed. For (\ref{Emequal}), calculations from \cite{sema15b} predict $\phi=1$, which indeed dramatically improves the approximate energies, but with a loss of the variational character. Results from \cite{sema19c} show that $Q_\phi(N)\,\hbar$ is a good estimation of $I$ for the eigenstates of $H$. The quantum Kepler invariant is then
\begin{equation}
\label{taumequal}
\tau_q = \frac{\pi}{4}\, G \, m^{5/2} N(N-1)^{3/2}.
\end{equation}
Let us remark that this result is independent from the value of $Q_\phi(N)$, as this quantity cancels out in the calculation. For this reason, (\ref{taumequal}) can possibly give a precise result, since the main source of inaccuracy in the ET calculation is the structure of the global quantum number \cite{sema15a,sema15b}. Moreover, as the quantum character of $I$ and $E$ is carried by the quantity $Q_\phi(N)\,\hbar$, this cancellation explains why the result~(\ref{taumequal}) could be relevant for a classical system. 

Classical and quantum Kepler invariants coincide for $N=2$ but differ for $N\ge 3$. More precisely, $\tau_q$ is $N(N-1)/2$ times the classical value given in \cite{sun18}. This discrepancy can be due to the fact that the classical Kepler invariant in  \cite{sun18} has been determined to be in agreement with the invariant found numerically for three-body classical systems, these calculations taking into account the free group element $L_f$ for which it is difficult to find a quantum equivalent. 

\begin{figure}[ht]
\includegraphics[width=8cm]{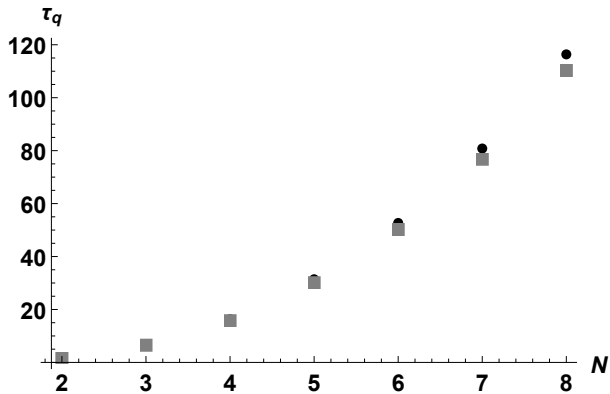}
\caption{Kepler invariant $\tau_q$ for the ground state of self-gravitating bosons ($\hbar = G = m =1$) as a function of $N$. Black circles: values of (\ref{taumequal}); Gray squares and diamonds: results from (\ref{tauq}) with $E$ taken from data in \cite{horn14} and $I=Q_\phi(N)\,\hbar$, respectively for $\phi=1$ and $\phi=2$.}\label{fig:redtime}
\end{figure}

Thanks to the accurate energies computed in \cite{horn14} for the ground state ($n_i = l_i = 0$, $\forall\ i$) of self gravitating bosons with $\hbar = G = m =1$, it is possible to check the relevance of the notion of quantum Kepler invariant on a particular example. As values of $I$ are not available from \cite{horn14}, values for $\tau_q$ computed with (\ref{taumequal}) and with (\ref{tauq}), under the assumption that $I=Q_\phi(N)\,\hbar$ , are compared in Fig.~\ref{fig:redtime}. One can see that the agreement is very good for $\phi=1$ (this value giving also the best agreement for the energies computed with the ET). The two formulas coincide for $N=2$, as already mentioned. This shows that the choice of $Q_1(N)\,\hbar$ for $I$ seems quite reasonable.

\section{Quantum systems with different particles}
\label{sec:gen}

The ET has been generalized to study systems with different particles (bosons or fermions) \cite{sema20}, but the procedure is then more complicated to implement. In order to well understand the approximations involved, let us detail a little bit the calculations. The first step of the procedure is to build an auxiliary Hamiltonian \cite{sema20}
\begin{equation}
\label{Htilde}
\tilde H = \sum_{i=1}^N \frac{\bm p_i^2}{2 m_i} + 
\sum_{i<j=2}^N \left[\rho_{ij}\,\bm r_{ij}^2 - \frac{3}{2^{2/3}}\left( G\,  m_i\,  m_j \right)^{2/3}\rho_{ij}^{1/3} \right].
\end{equation}
The upper bounds of the eigenvalues of $H$ are then determined by minimizing the eigenvalues $\tilde E\left( \{ \rho_{ij} \}\right)$ of $\tilde H$ with respect to the auxiliary parameters $\{ \rho_{ij} \}$. Unfortunately, the values $\tilde E$ cannot be analytically computed in general when $N > 5$ \cite{silv10,hall79,ma00}. As our purpose is to obtain general analytical results, we must resort to a supplementary approximation. Slightly worse upper bounds will be obtained if we impose the constraint 
\begin{equation}
\label{rho}
\rho_{ij}=\rho\, m_i\, m_j. 
\end{equation}
The eigenvalues $\tilde E$ can then be computed using the procedure detailed in \cite{silv10,hall79,ma00} 
\begin{equation}
\label{Etilde}
\tilde E(\rho) = \sqrt{2\rho\sum_{i=1}^N m_i}\,Q_\phi(N)\,\hbar-\frac{3}{2^{2/3}}G^{2/3}\rho^{1/3} \sum_{i<j=2}^N m_i\, m_j.
\end{equation}
Equation (\ref{Etilde}) is exact only for $\phi =2$, but we take the liberty to replace $Q_2(N)$ by $Q_\phi(N)$. The minimization with respect to $\rho$ gives upper bounds of the upper bounds
\begin{equation}
\label{Etilde0}
\tilde E = - \frac{G^2}{2\, Q_\phi(N)^2\,\hbar^2} \frac{\left(\sum_{i<j=2}^N m_i\,  m_j\right)^3}{\sum_{i=1}^N m_i}.
\end{equation}
This result coincides exactly with the formula (10) in \cite{sun19}, guessed solely on the basis of (\ref{tauqgen}) below. If we assume again that $Q_\phi(N)\,\hbar$ is a good estimation of $I$, (\ref{tauq}) gives 
\begin{equation}
\label{tauqgen}
\tau_q = \frac{\pi}{\sqrt{2}}\, G \left[\frac{\left(\sum_{i<j=2}^N m_i\,  m_j\right)^3}{\sum_{i=1}^N m_i}\right]^{1/2},
\end{equation}
which is the relation (2) in \cite{sun19}, determined on the basis of dimensional arguments. Let us remark that this result is independent from the value of $\phi$. So, at this level of approximation, an universal Kepler invariant is also obtained for general systems. It is easy to see that (\ref{tauqgen}) is equal to the classical equivalent (\ref{kepler3L}) when $N=2$, and that (\ref{taumequal}) is recovered when $m_i=m, \ \forall\ i$.

\begin{figure}[ht]
\includegraphics[width=8cm]{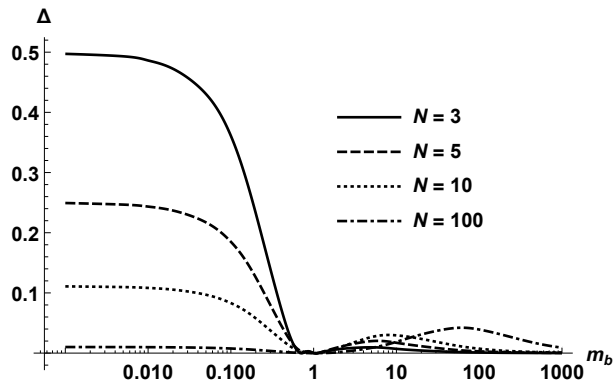}
\caption{Relative error $\Delta$ with $\phi=2$ (see text) for bosonic ground states as a function of $m_b$ with $\hbar = G = m_a =1$, for some values of $N$.}\label{fig:Deltatau}
\end{figure}

In order to check the error made with the approximation~(\ref{rho}), let us consider a system composed of $(N-1)$ particles with a mass $m_a$ and the last one with a mass $m_b$. In this case, $\rho_{ij}=\rho_{aa}$ for $i\le j < N$ and $\rho_{iN}=\rho_{ab}$, and $\tilde H$ can be solved \cite{sema20}
\begin{align}
\label{EtildeNa}
\tilde E(\rho_{aa},\rho_{ab}) &= \sqrt{\frac{2\left((N-1)\,\rho_{aa}+\rho_{ab}\right)}{m_a}} \,Q_\phi(N-1)\,\hbar + \sqrt{\frac{2\left((N-1)\,m_a+m_b\right)\rho_{ab}}{m_a m_b}}\,Q_\phi(2)\,\hbar \nonumber \\
&-\frac{3\, (N-1)(N-2)}{2^{5/3}} (G\, m_a^2)^{2/3}\rho_{aa}^{1/3}  -\frac{3\, (N-1)}{2^{2/3}} (G\, m_a\, m_b)^{2/3}\rho_{ab}^{1/3}
\end{align}
The upper bound of the ET can be numerically obtained by computing $\tilde E =\min_{\rho_{aa},\rho_{ab}} \tilde E(\rho_{aa},\rho_{ab})$. The Kepler invariant $\tilde \tau_q$ is calculated with this value $\tilde E$ and $I=Q_\phi(N)\,\hbar$, and compared with the value $\tau_q$ computed with (\ref{tauqgen}) by using the relative error $\Delta$ defined by
\begin{equation}
\label{Delta}
\Delta = \frac{\tilde \tau_q- \tau_q}{\tilde \tau_q}.
\end{equation}
This error is presented in Fig.~\ref{fig:Deltatau} for bosonic ground states ($n_i = l_i = 0$, $\forall\ i$) with $\phi=2$, but calculations show that the results are practically independent from $\phi$. The cancellation of $\phi$ effects seems nearly perfect in this case. It is remarkable that the error due to the calculation of $\tau_q$ with (\ref{tauqgen}) is quite small, except when the number of particles is small and the different particle is much lighter than the other ones.  

Let us remark a particular point. The ET solutions for $N$ identical bosons are defined for completely symmetrized states \cite{silv10}. For the system considered here, with $N_a$ particles plus a different one, the ET solutions are defined for states symmetrized only for the $N_a$ first particles \cite{sema20}. If the last particle is set identical to the other ones (same mass and same interaction), one could expect solutions which are different from the system with all identical bosons. But the ET is such that this is not the case. That is why $\Delta=0$ for $m_b = m_a = 1$ on Fig.~\ref{fig:Deltatau}. This situation has also been noted for other bosonic systems studied in \cite{sema20} but has not yet received a general demonstration.

\section{Concluding remarks}
\label{sec:conclu}

Our calculations give strong indications in favor of the existence of a quasi Kepler's third law for quantum self-gravitating particles. The envelope theory used to approximately solve the quantum many-body problem predicts an exact Kepler invariant for systems with identical particles, and a quasi exact invariant for systems with different particles. For identical particles, a check is performed thanks to numerical data available in \cite{horn14} for the ground states of bosons. For different particles, the relevance of the approximate invariant obtained is only verified for systems with a set of identical particles plus a different one. It is certainly desirable for more types of systems to be studied. Above all, accurate numerical calculations like the ones achieved in \cite{horn14} should be extended to systems with different particles. 

The invariant obtained in the general quantum case is different from the one computed for the classical equivalent system \cite{sun18,sun19}. This is probably due to the introduction in the classical calculations of information about the topology of the classical orbits, for which no equivalent is found in quantum calculations. Nevertheless, if a quasi Kepler's third law exists for quantum self-gravitating particles, it is worth considering its existence for equivalent many-body classical systems. So, supplementary studies like \cite{dmit15,li17,li18,li19} are certainly desirable for classical orbits with more than three particles.

Let us mention that an invariant also seems to exist for collisionless periodic orbits in the Coulomb potential for three charged 
particle, one positive and two negative \cite{sind18}. The presence of attractive and repulsive interactions in this system makes it different from the purely attractive cases presented above. So further research involving various Coulombic systems seem desirable to check the existence of some universal relation for such systems.

If these quasi Kepler's third laws for quantum and classical many-body systems are something else than happy coincidences, it is worth searching for some fundamental principle at work. This problem certainly deserves further research.

\begin{acknowledgments}
This work was supported by the Fonds de la Recherche Scientifique - FNRS under Grant Number 4.45.10.08. 
\end{acknowledgments}

\end{document}